\documentclass[a4paper,11pt]{article}
\usepackage{pos}

\DeclareMathOperator{\Tr}{Tr}

\title{The temperature dependence of fractional topological charge objects}

\author*[a]{Jackson A. Mickley}
\author[a]{Waseem Kamleh}
\author[a]{Derek B. Leinweber}

\affiliation[a]{Centre for the Subatomic Structure of Matter, Department of Physics, \\
	The University of Adelaide, South Australia 5005, Australia}

\emailAdd{jackson.mickley@adelaide.edu.au}
\emailAdd{waseem.kamleh@adelaide.edu.au}
\emailAdd{derek.leinweber@adelaide.edu.au}

\abstract{We present a novel method for defining the topological charge contained within distinct topological objects in the nontrivial ground-state fields of $\mathrm{SU}(N)$ lattice gauge theory. Such an analysis has been called for by the growing number of models for Yang-Mills topological structure which propose the existence of fractionally charged objects. This investigation is performed for $\mathrm{SU}(3)$ at a range of temperatures across the deconfinement phase transition, providing an assessment of how the topological structure evolves with temperature. This reveals a connection between the topological charge and holonomy of the system which must be satisfied by finite-temperature models of Yang-Mills vacuum structure. We find a promising consistency with the instanton-dyon model for $\mathrm{SU}(N)$ vacuum structure.}

\FullConference{The XVIth Quark Confinement and the Hadron Spectrum Conference (QCHSC24)\\
 19-24 August, 2024\\
 Cairns Convention Centre, Cairns, Queensland, Australia\\}

\begin{document}
\maketitle

\section{Introduction}
There is now a wealth of literature revealing that fractional-topological-charge solutions to the Yang-Mills equations play a vital role in $\mathrm{SU}(N)$ vacuum structure. The earliest such constructions date back to the early 1980s \cite{tHooft:1981nnx} and have been extensively studied in numerical simulations \cite{Gonzalez-Arroyo:1995isl, Gonzalez-Arroyo:1995ynx, GarciaPerez:1989gt, GarciaPerez:1992fj, GarciaPerez:1997fq, Montero:2000mv}. Such ``fractional instantons'' possess $\mathbb{Z}_N$ flux and are thus a possible microscopic mechanism for confinement \cite{Gonzalez-Arroyo:1995isl, Gonzalez-Arroyo:1995ynx, Gonzalez-Arroyo:2023kqv}. However, the breakthrough came with the discovery of the caloron \cite{Lee:1998vu, Lee:1998bb, Kraan:1998sn, Kraan:1998pm} which generalises the instanton to finite temperature. The caloron profile can be viewed as composed of $N$ constituents, each possessing fractional charge depending on the Polyakov loop at spatial infinity. This provides an intimate connection to confinement. Models of the finite-temperature Yang-Mills vacuum in terms of these ``instanton-dyons'' \cite{Diakonov:2007nv, Diakonov:2010qg, Shuryak:2013tka, Liu:2015ufa} have successfully reproduced the second-order phase transition of $\mathrm{SU}(2)$ \cite{Larsen:2015vaa, Lopez-Ruiz:2016bjl, Lopez-Ruiz:2019oov} and the first-order phase transition of $\mathrm{SU}(3)$ \cite{DeMartini:2021dfi}.

In this work, we devise a novel algorithm for estimating the topological charge contained within distinct topological objects in $\mathrm{SU}(N)$ lattice gauge theory. This is employed for $\mathrm{SU}(3)$ at a range of temperatures across the phase transition to provide an assessment of how the topological structure evolves with temperature. The findings reveal an inherent relation between topological structure and confinement, eliciting a connection to the instanton-dyon model.

\section{Method}
We seek to identify and calculate the charge contained within the distinct topological objects present in the Yang-Mills vacuum. This is achieved by identifying objects through their peaks in the topological charge density and then iteratively assigning lattice sites to each object in turn, effectively ``growing'' the objects. Subject to a few reasonable assumptions, we can be confident the set of points assigned to an object reflects its distribution of topological charge density. The charge of each individual object can then be calculated by summing the topological charge density over their allocated sets of points.

Denoting the topological charge density by $q(x)$, the algorithm proceeds as follows:
\begin{enumerate}
	\item Identification: identify the local maxima for $q(x) > 0$, and minima for $q(x) < 0$, present in $q(x)$ within a $3^4$ hypercube. These are taken as the peaks of each distinct topological object.
	
	\item Growing: first, consider the object with the smallest peak and take the set of points currently assigned to that object. Assign all neighbouring points that satisfy the below criteria to the same object.
	\begin{itemize}
		\item Have not already been assigned to an object, such that information is not overwritten.
		\item Have the same sign topological charge density, such that positive and negative topological excitations are kept distinct.
		\item Have a lower (absolute) topological charge density value, such that we only grow downhill.
	\end{itemize}
	This is repeated for all objects in order of increasing peak size. Lower-peaked objects are prioritised as edge points have a greater relative weight compared to sharply peaked objects.
	
	\item Iteration: repeat the Growing process (Step 2) until all valid lattice sites have been assigned.
	
	\item Dislocation filtering: filter out the peaks that fail to attain all points within the surrounding $3^4$ hypercube. These are considered lattice dislocations instead of genuine topological objects.
	
	\item Finishing up: reinstate the Iteration (Step 3) for the surviving objects, thereby reallocating the newly available sites.
\end{enumerate}

\begin{figure}
	\centering
	\includegraphics[width=0.37\linewidth]{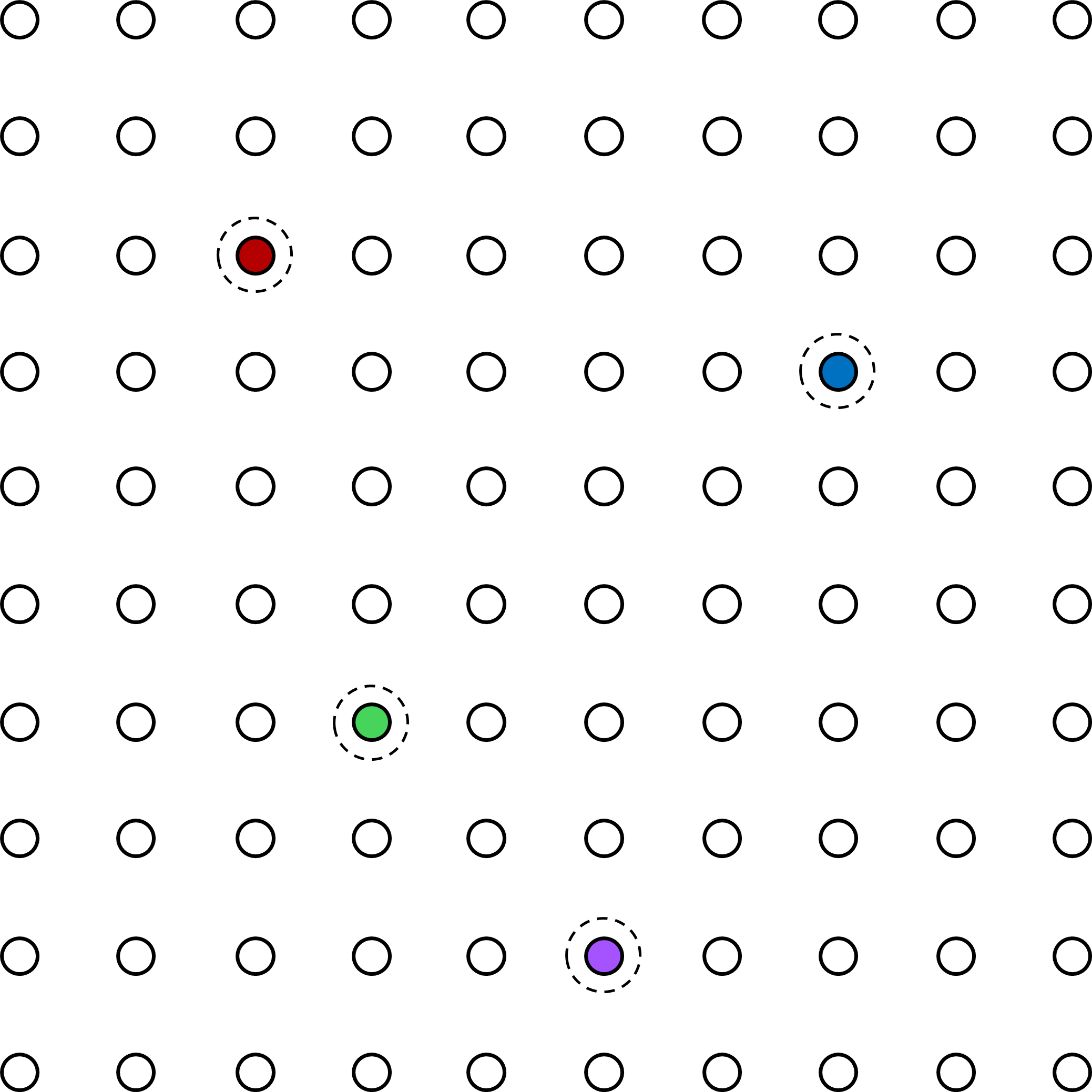}
	\hspace{4em}
	\includegraphics[width=0.37\linewidth]{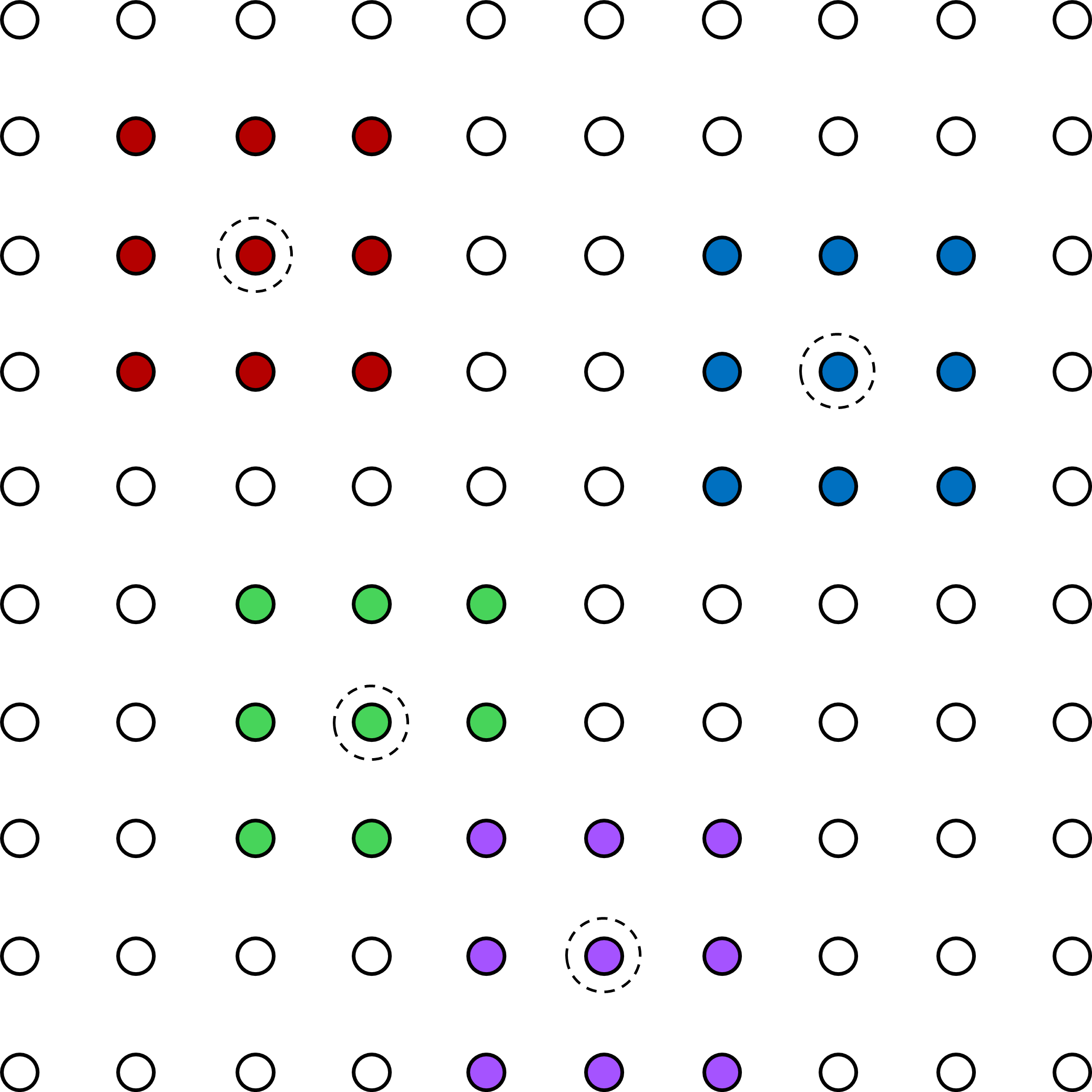}
	
	\vspace{3em}
	
	\includegraphics[width=0.37\linewidth]{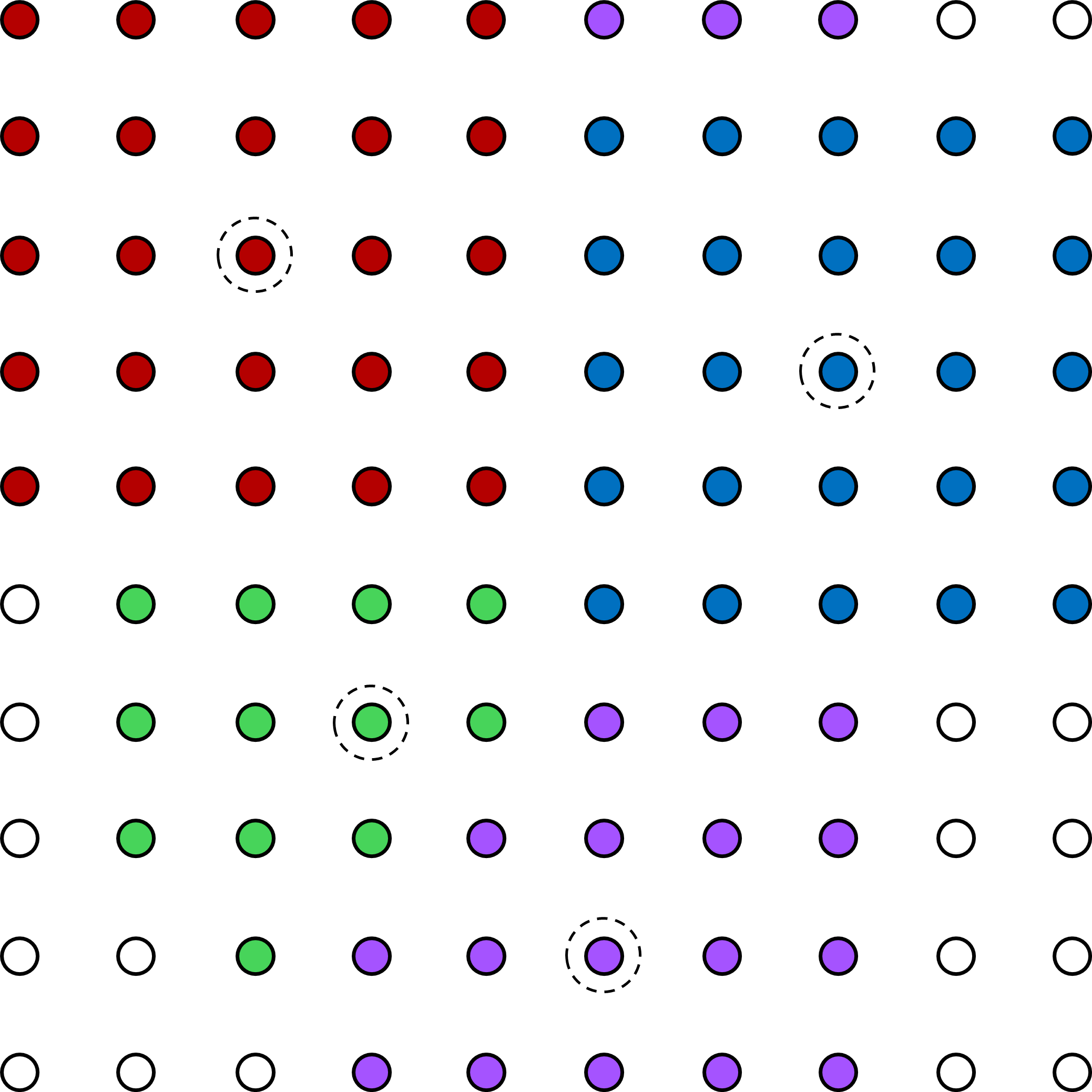}
	\hspace{4em}
	\includegraphics[width=0.37\linewidth]{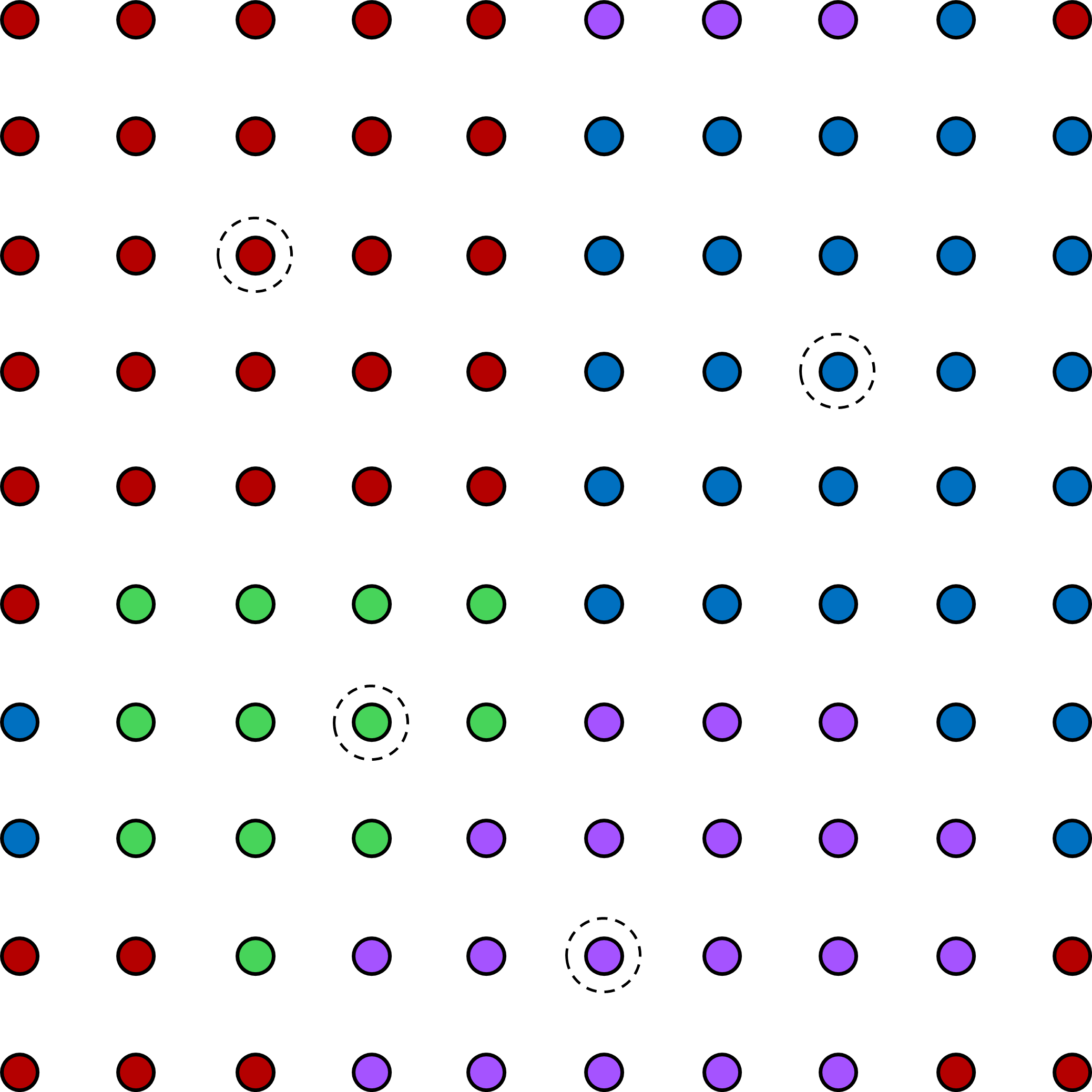}
	
	\vspace{3em}
	
	\includegraphics[width=0.37\linewidth]{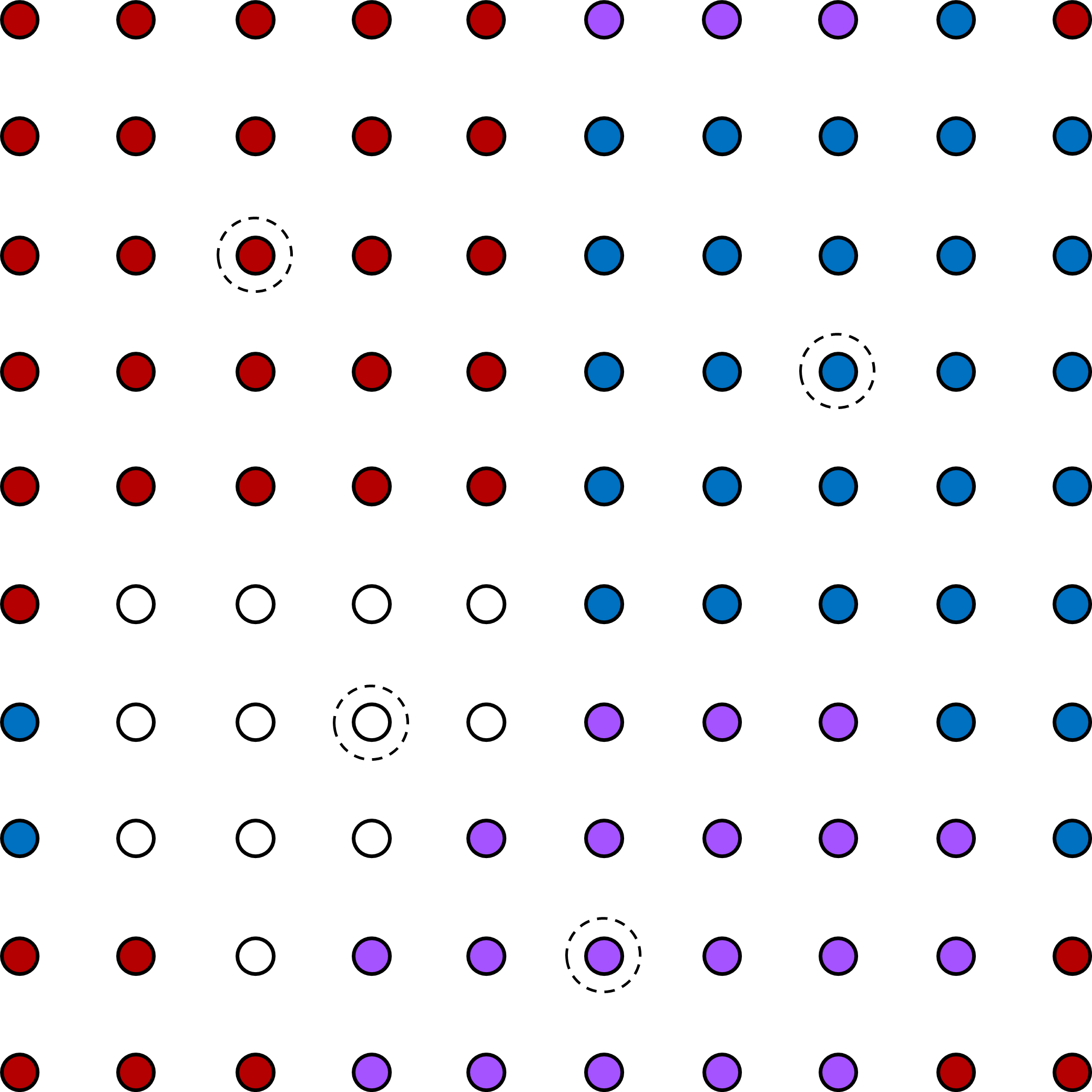}
	\hspace{4em}
	\includegraphics[width=0.37\linewidth]{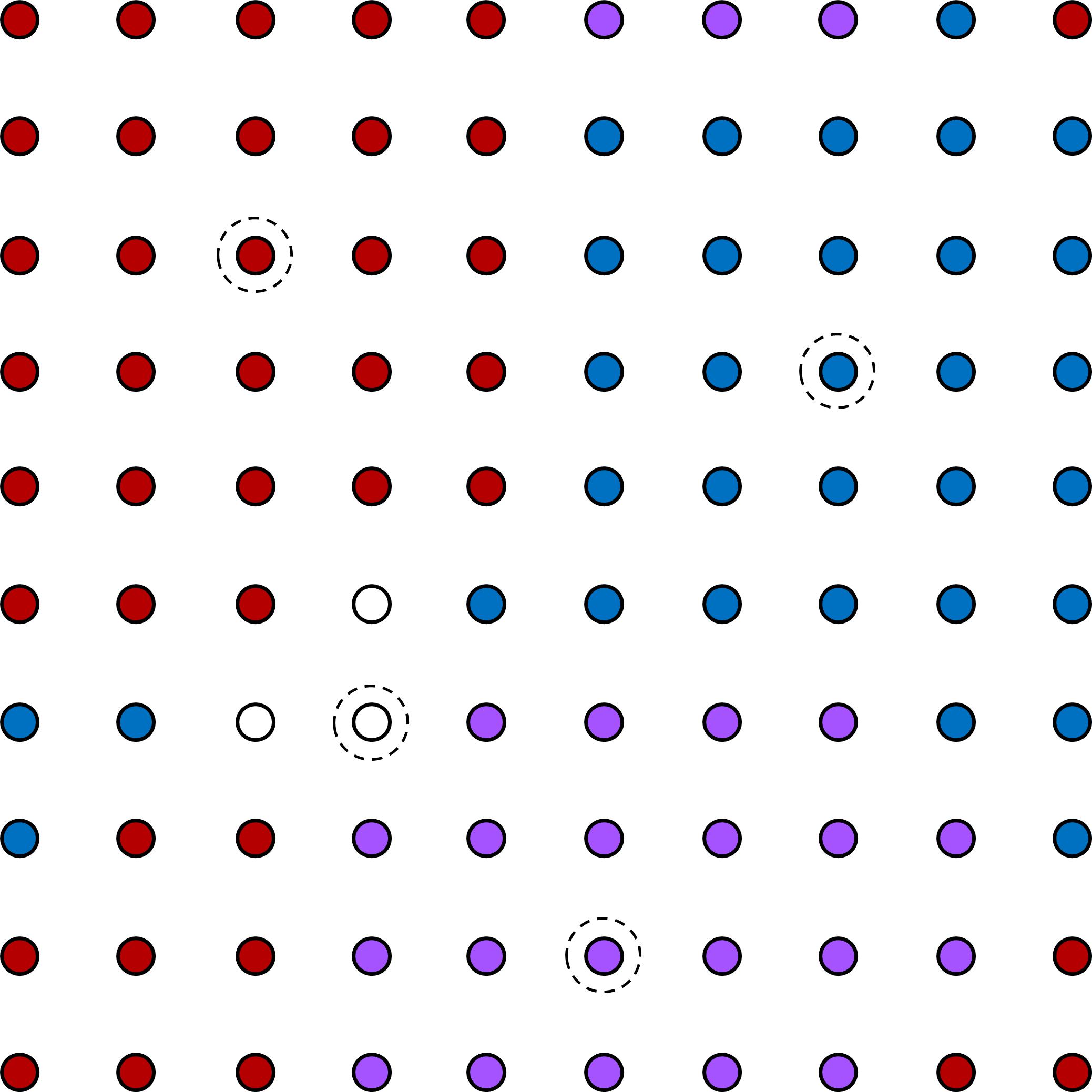}
	
	\vspace{0.4em}
	
	\caption{\label{fig:algorithm} A graphic demonstrating our algorithm, proceeding from left to right then top to bottom. The order in which objects grow is red $\rightarrow$ blue $\rightarrow$ purple $\rightarrow$ green. First, the peaks are identified and neighbouring hypercubes assigned (\textbf{top row}). The objects are then allowed to grow until all possible points have been assigned (\textbf{middle row}). Finally, we discard the peaks that fail to assign all points within the $3^4$ hypercube, and the remaining objects grow until no more points satisfy the required criteria (\textbf{bottom row}). In this case, these are the green points.}
\end{figure}

At this point, each physical object (as defined by the algorithm) has a set of points allocated to it that allows a straightforward computation of their topological charges. The above process is illustrated visually in Fig.~\ref{fig:algorithm}.

\begin{figure}
	\centering
	\includegraphics[width=0.6\linewidth]{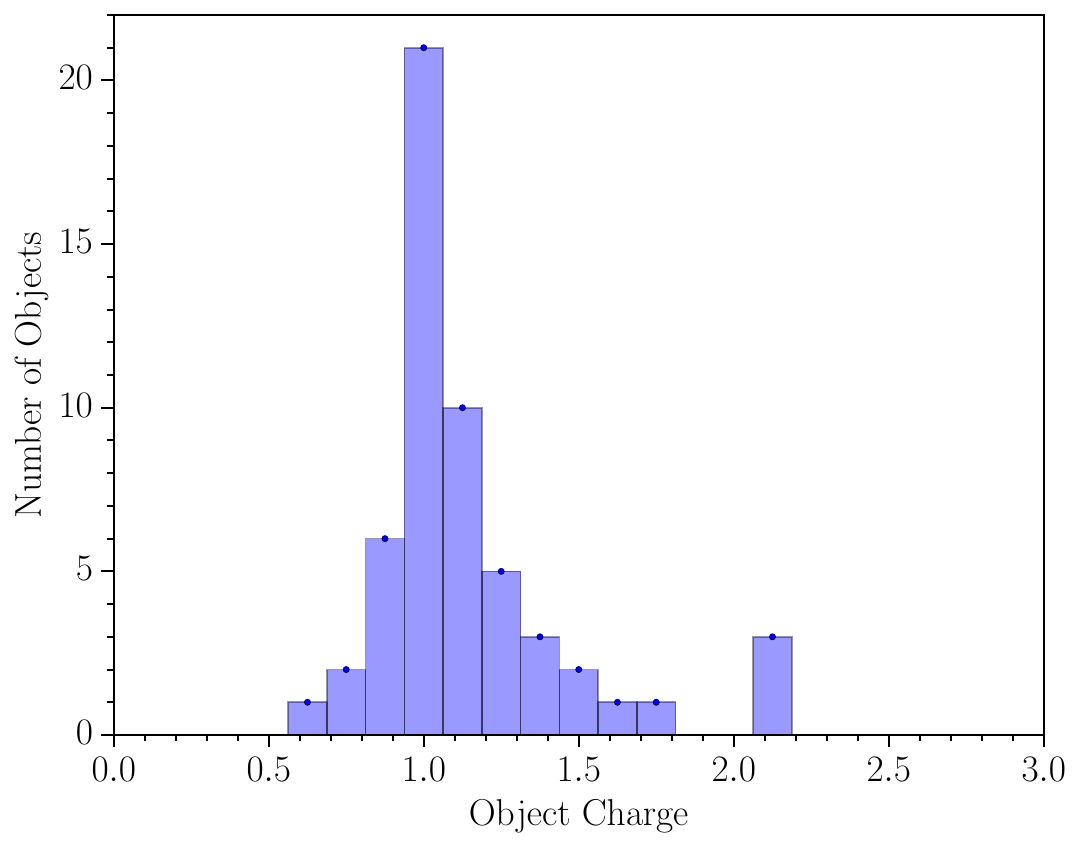}
	\caption{\label{fig:classicallimit} The result of applying our algorithm to five $32^3\times 64$ configurations after 4000 sweeps of $\mathcal{O}(a^4)$-improved cooling.}
\end{figure}

The algorithm can be tested by applying it to configurations that have undergone extended cooling \cite{Berg:1981nw, Teper:1985rb, Ilgenfritz:1985dz}, which approaches the classical limit. Here, we expect to find exclusively instantons with integer topological charge \cite{Teper:1985rb, Ilgenfritz:1985dz}. The result of applying the algorithm in this limit is shown in Fig.~\ref{fig:classicallimit} as a histogram.
This can be read as giving the number of identified objects with a calculated topological charge within some narrow interval. We find that the histogram is strongly peaked near 1, with a small spread of values slightly farther away. This distribution of calculated topological charges can be understood by visualising the topological charge density.

\begin{figure*}
	\centering
	\includegraphics[width=0.44\linewidth]{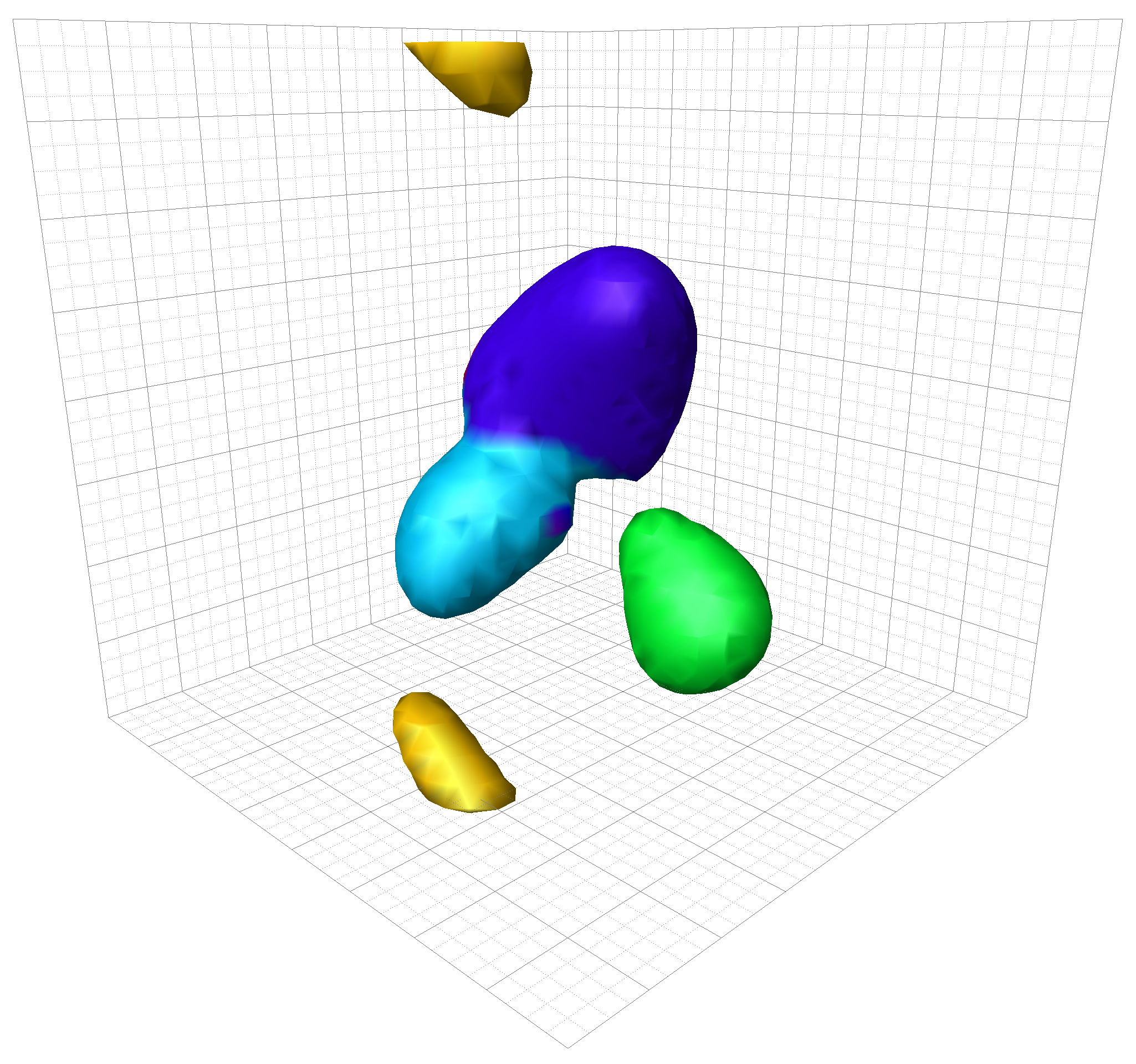}
	\hspace{2em}
	\includegraphics[width=0.44\linewidth]{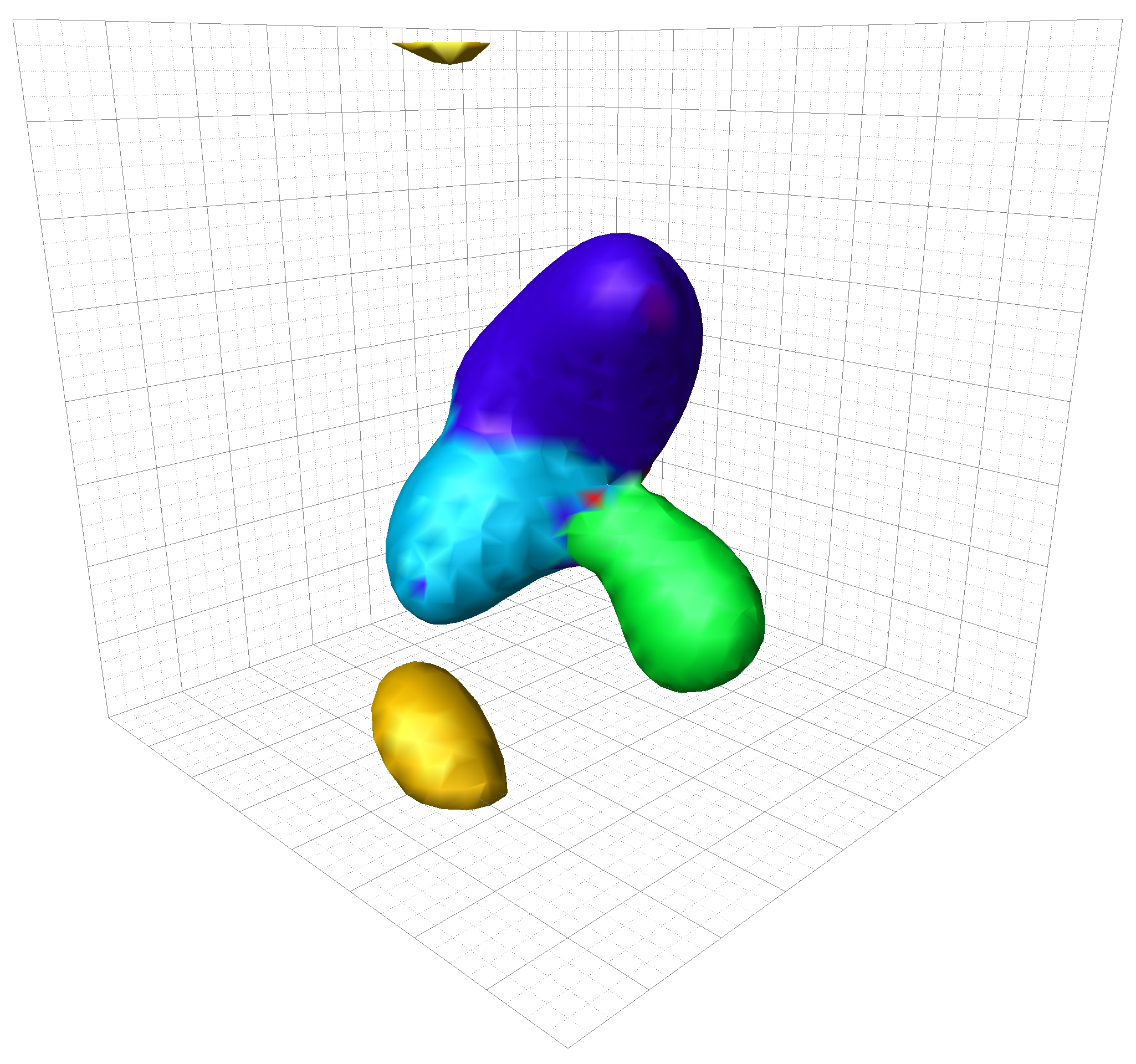}
	
	\vspace{0.5em}
	
	\includegraphics[width=0.44\linewidth]{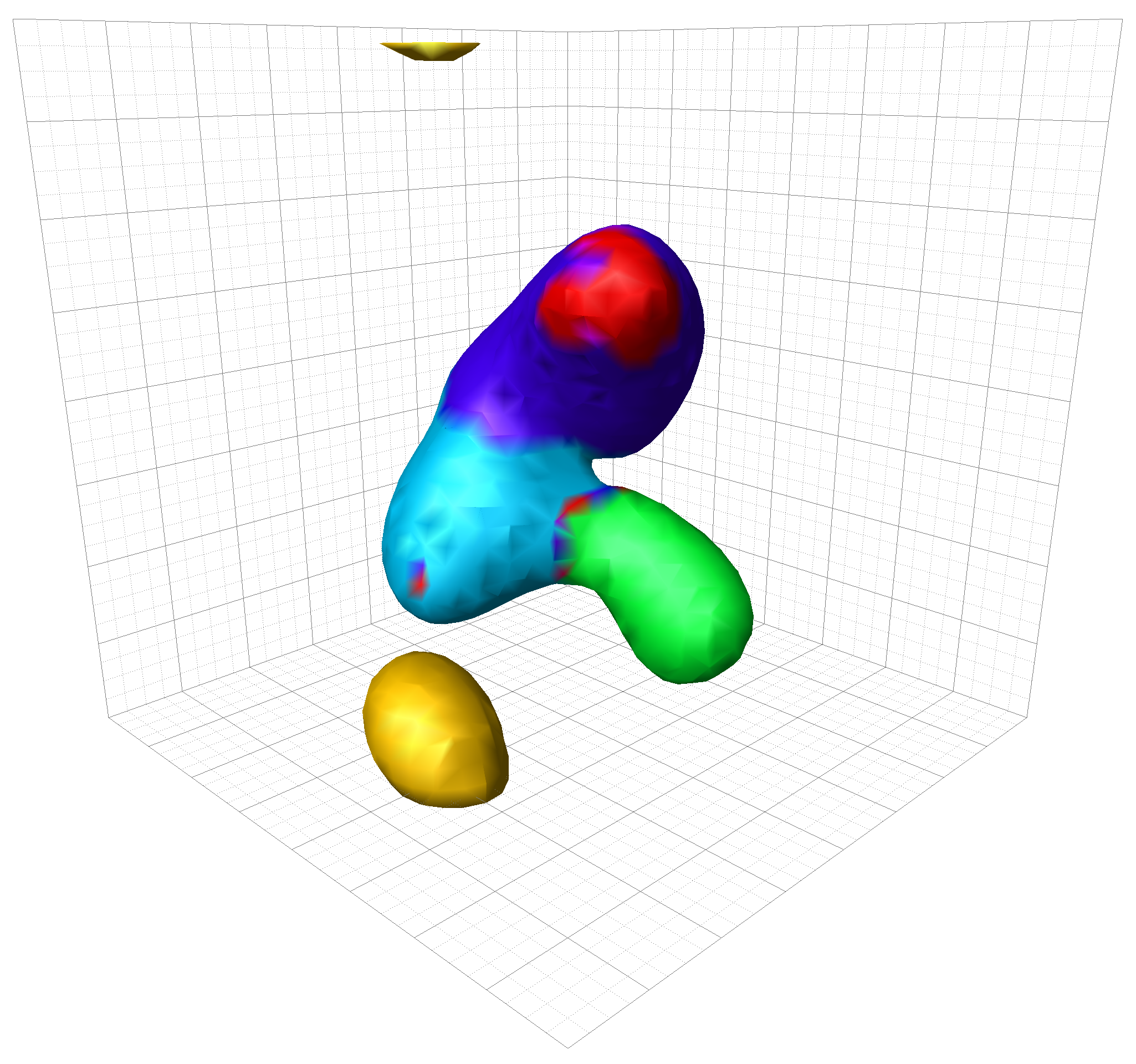}
	\hspace{3em}
	\includegraphics[width=0.44\linewidth]{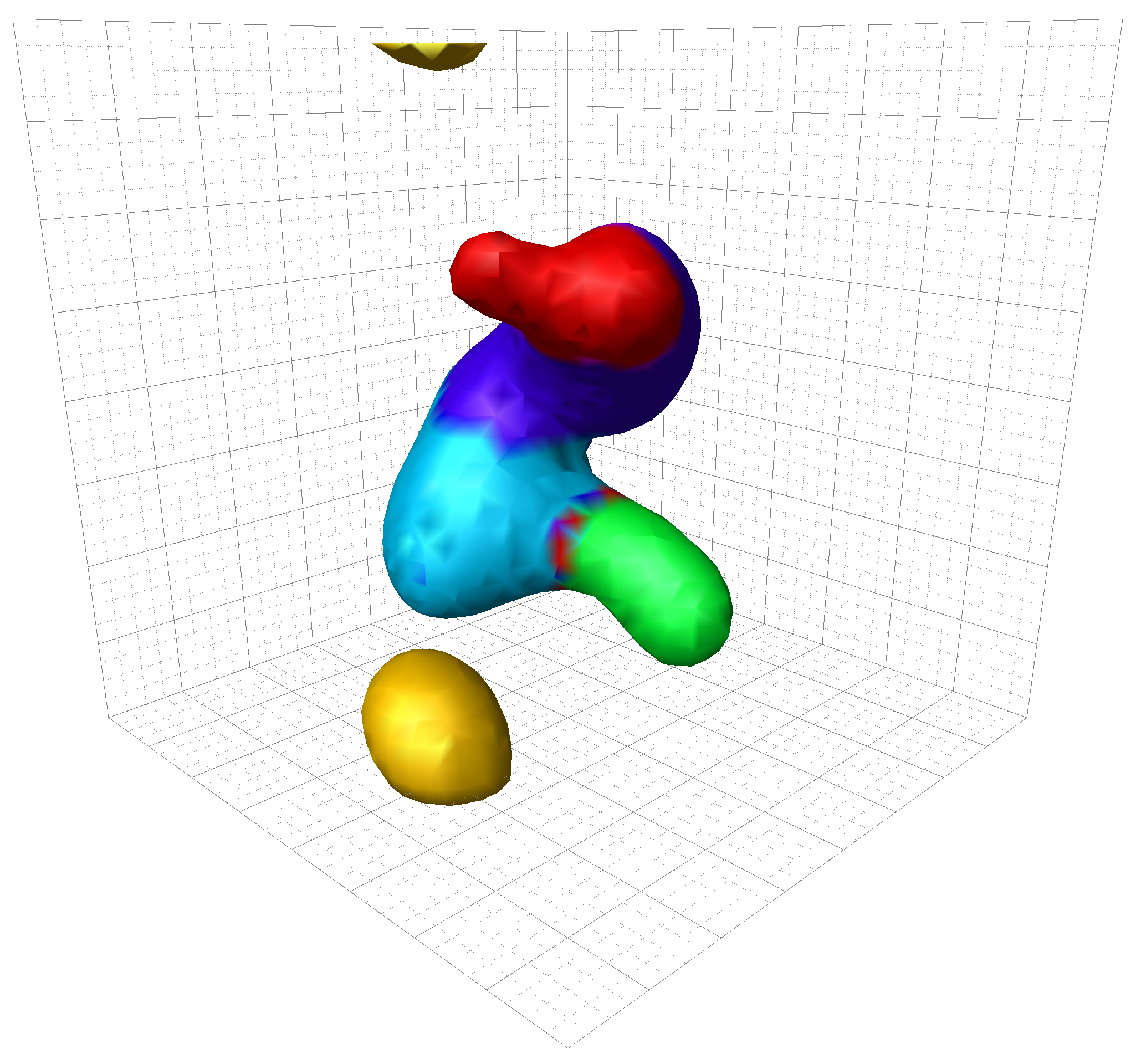}
	
	\vspace{0.5em}
	
	\includegraphics[width=0.44\linewidth]{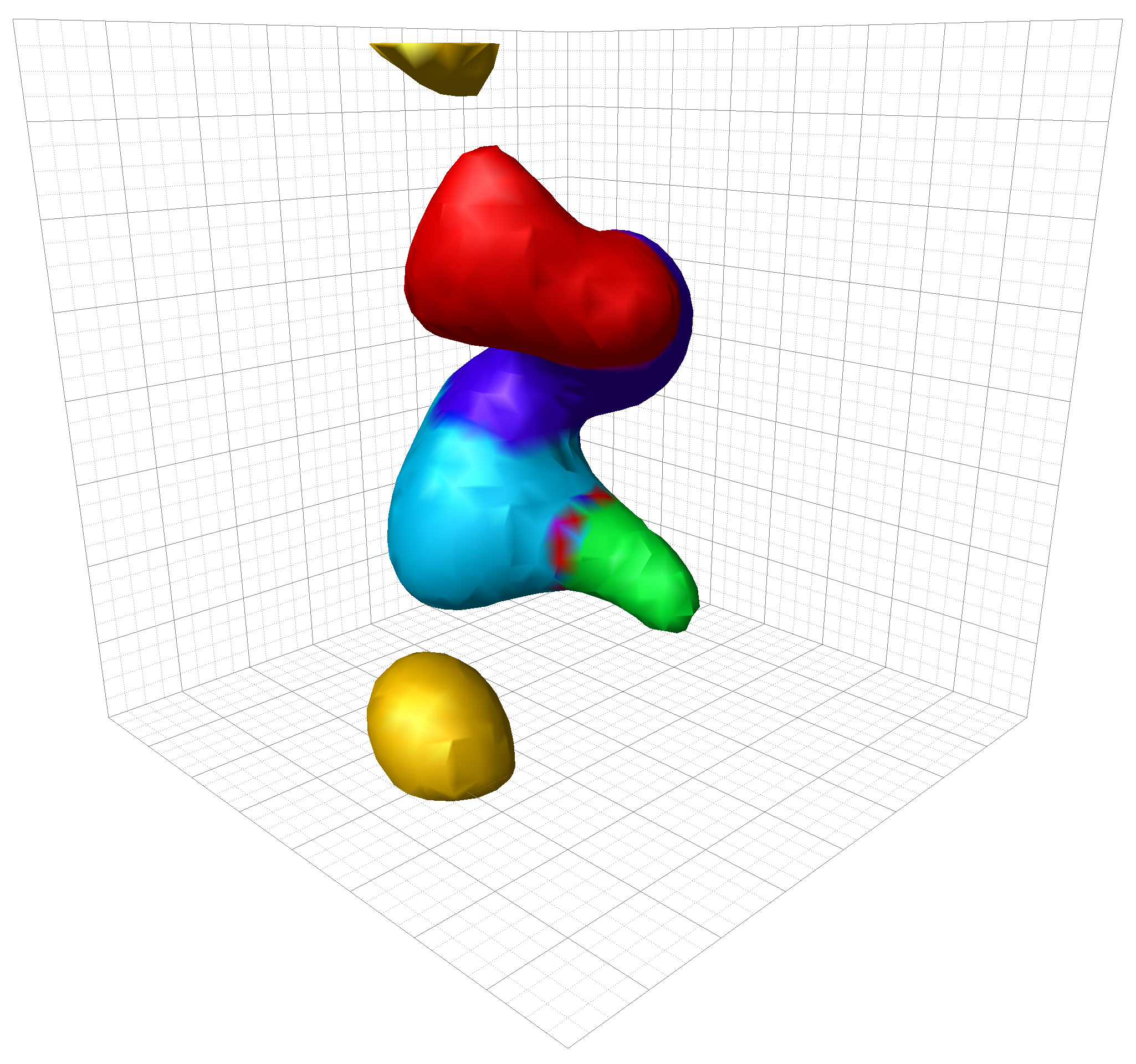}
	\hspace{2em}
	\includegraphics[width=0.44\linewidth]{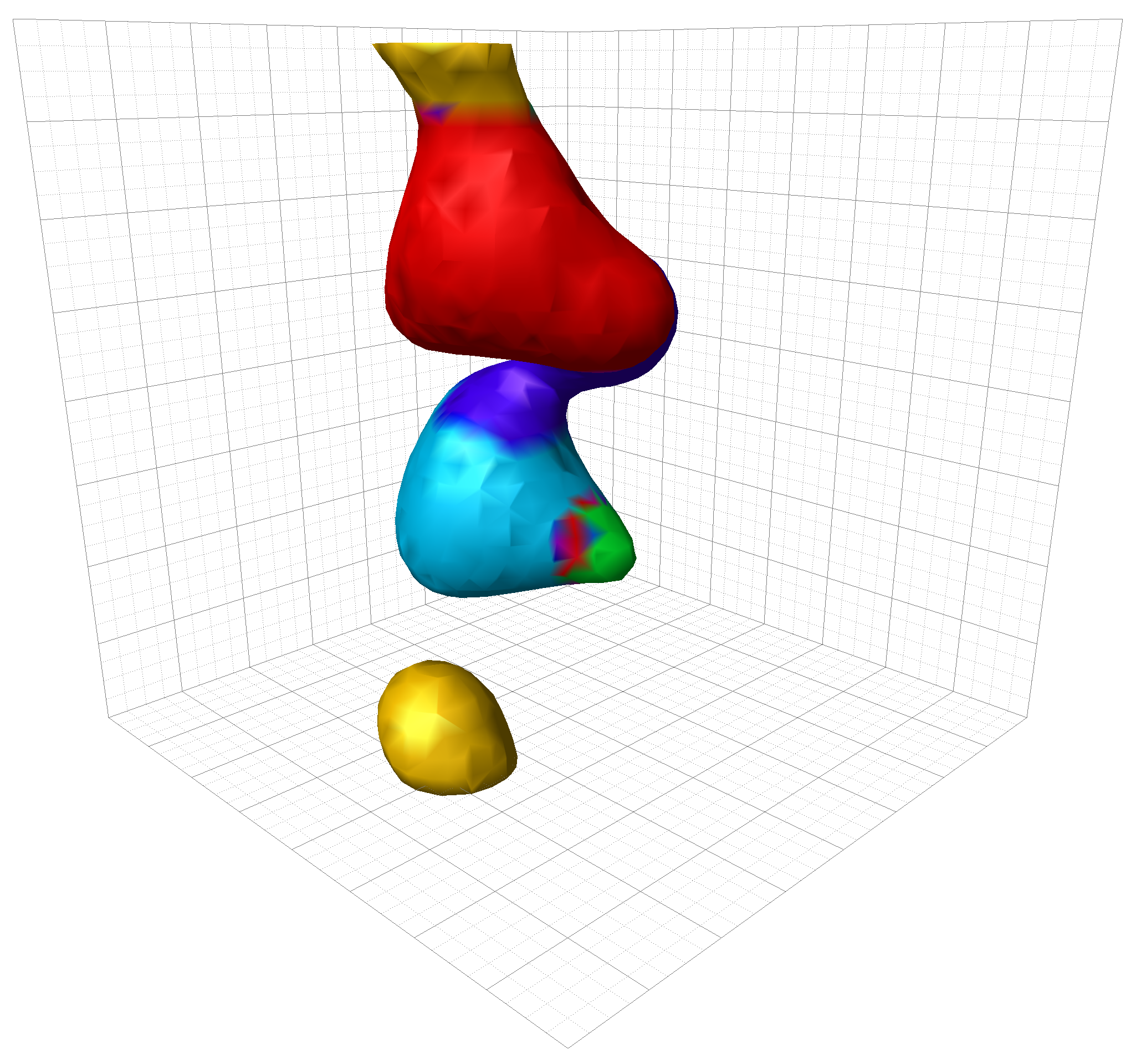}
	\caption{\label{fig:classicalvis} Visualisations of the topological charge density under extended cooling in three-dimensional slices of the full four-dimensional spacetime. There are six consecutive slices, displayed from left to right then top to bottom. The topological charge density is coloured according to the object number in the algorithm, allowing insight into how it divides topological objects in four dimensions. We only visualise the topological charge density above some minimum threshold value to observe the behaviour of the algorithm on the most significant topological charge density; this also enables one to see into the three-dimensional space.}
\end{figure*}

In Fig.~\ref{fig:classicalvis}, six consecutive three-dimensional slices of an example topological charge density under extended cooling are provided.
These visualisations are coloured according to the objects identified by the algorithm. This demonstrates the challenging four-dimensional nature of the problem even in this ``classical limit'', including how distinct objects are still strongly overlapping. For instance, the red object grows on top of the purple object as we advance in the fourth dimension, and subsequently merges with the yellow object.

As such, we do not expect to obtain perfect integers. This is due to fluctuations about the single-instanton solution, overlapping distributions and inherent uncertainties in dividing up the topological charge density in four dimensions, leading to systematics in how the lattice sites are allocated to objects. A histogram therefore provides a natural way to display the results.

Thus, the sharp peak in the topological-charge histogram (Fig.~\ref{fig:classicallimit}) with a narrow distribution is the expected outcome, and reflects the success of the algorithm in capturing the topological structure. A small peak near a topological charge of $2$ also signifies a possible contribution from strongly overlapping instantons \cite{GarciaPerez:1999zk, GarciaPerez:2000uk} and multi-instanton objects \cite{Witten:1976ck, Atiyah:1978ri}. In all, this finding provides strong evidence that the histogram mode obtained from a general configuration gives a reliable indicator of the underlying topological structure.

\section{Smoothing}
To probe the semiclassical structure of the Yang-Mills vacuum, a degree of UV-smoothing must be applied to tame the significant renormalisation factors encountered by naive lattice operators for the action and topological charge densities. We employ an $\mathcal{O}(a^2)$-improved plaquette $+$ rectangle gradient flow \cite{Luscher:2009eq, Luscher:2010iy} with step size $\rho = 0.005$.

Choosing the flow time at which to analyse the configurations is a nontrivial matter. We solve this by utilising two different $\mathcal{O}(a^4)$-improved topological charge density operators that experience substantially different renormalisation effects \cite{deForcrand:1997esx, Bilson-Thompson:2002xlt, Mickley:2023exg}. In one topological charge operator, we improve the field strength tensor $F_{\mu\nu}$ and square it to calculate the topological charge density. In the other, we directly improve the topological charge $q(x)$ analogous to action improvement schemes. Applying our algorithm to each of these operators in turn produces considerably different values for the topological objects' charges, if insufficient smearing has been performed. The primary way this manifests is through divergent histogram modes. This means we are unable to draw the same conclusions from both operators, even though they are each a valid discretisation of the topological charge density.

This gives a clear plan of action. We choose the minimum flow time required such that the two histogram modes are consistent with each other, allowing the same conclusions on topological structure to be drawn from each without oversmearing. This flow time of $\tau = 1.45$ is then kept constant with temperature, allowing an accurate comparison across all temperatures. In our results, we display both histograms overlaid to demonstrate this condition has been met.

\section{Topological structure}
\begin{figure}
	\centering
	\includegraphics[width=0.49\linewidth]{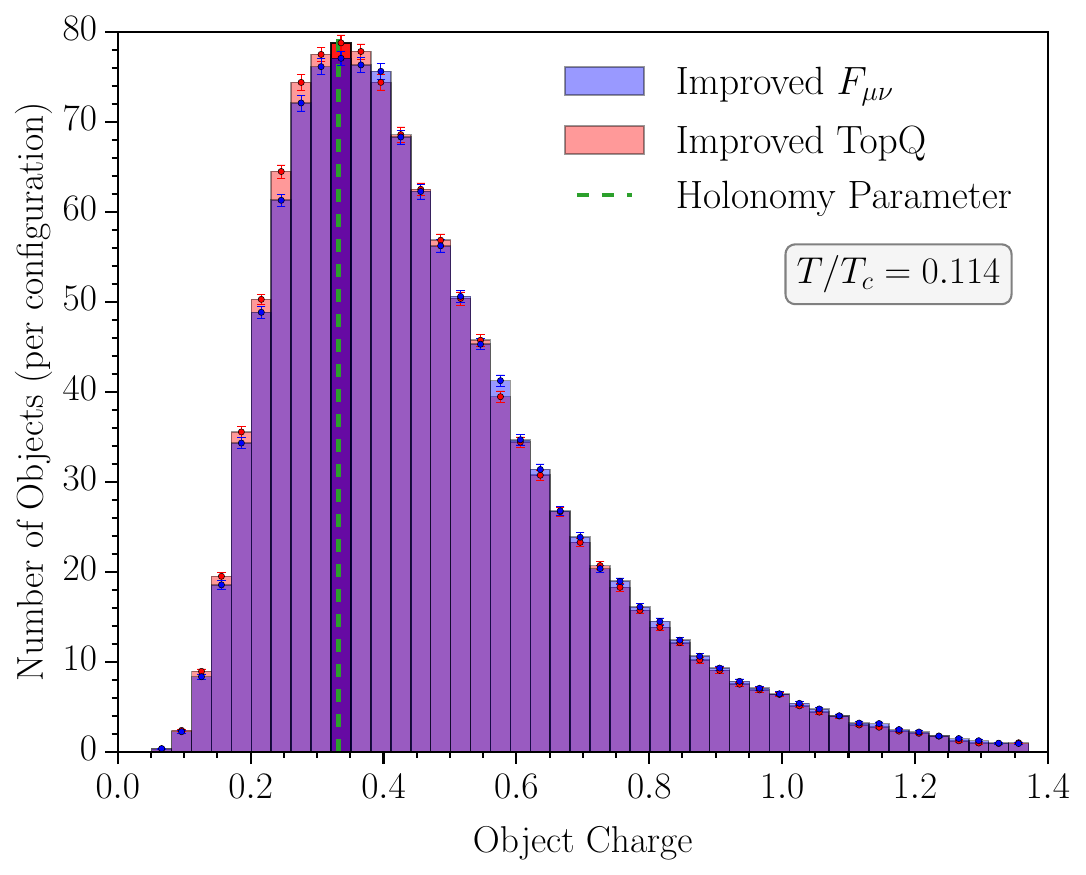}
	\hfill
	\includegraphics[width=0.49\linewidth]{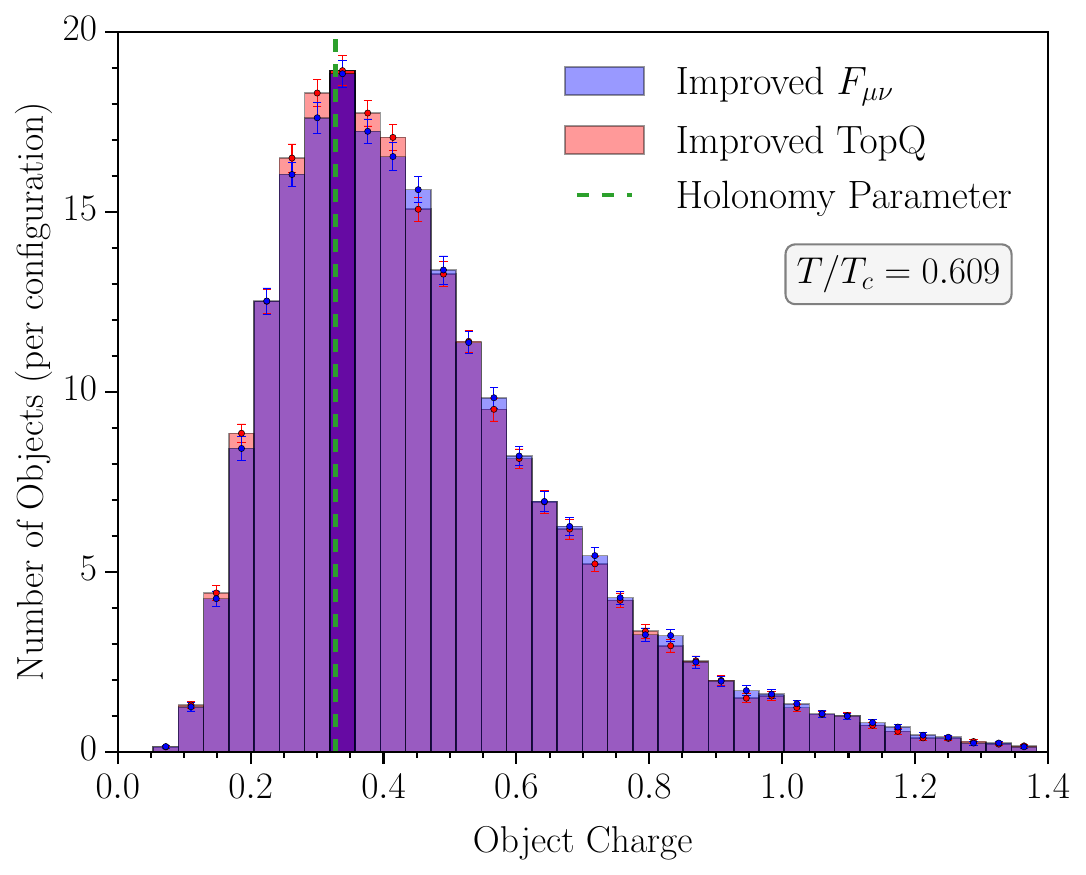}
	
	\vspace{0.5em}
	
	\includegraphics[width=0.49\linewidth]{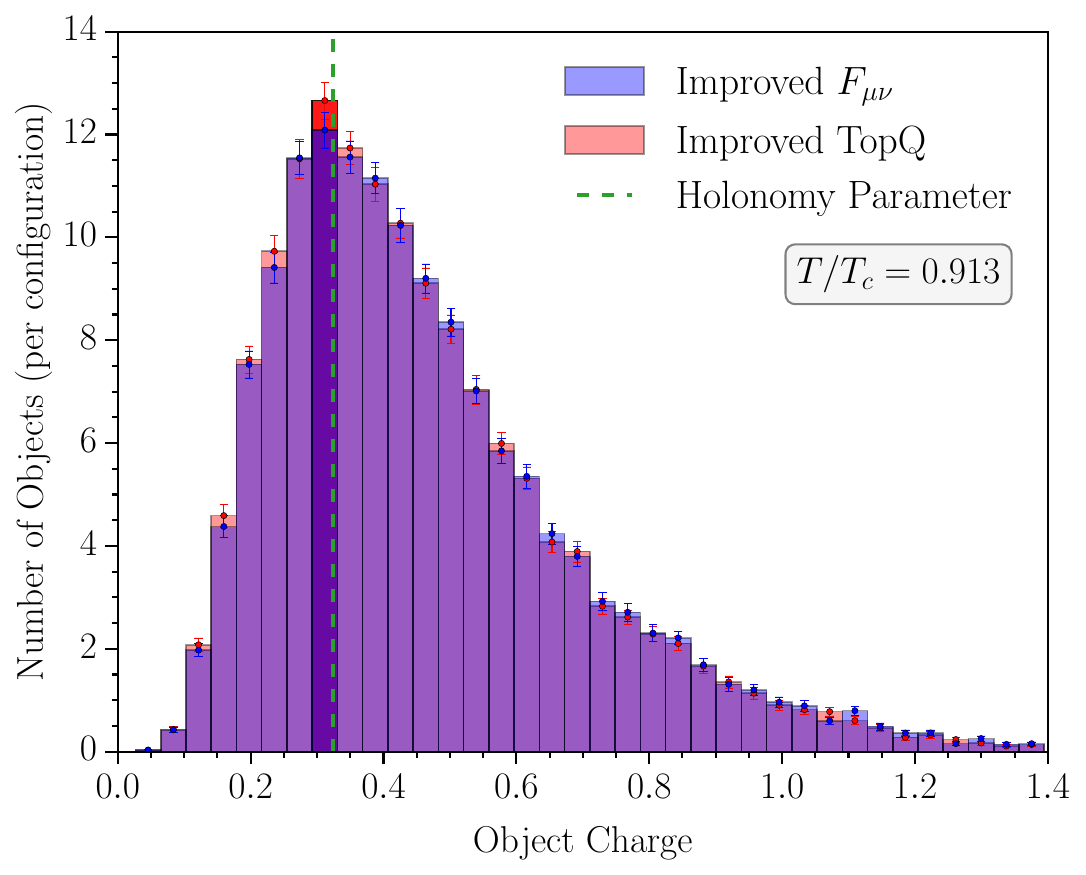}
	\hfill
	\includegraphics[width=0.49\linewidth]{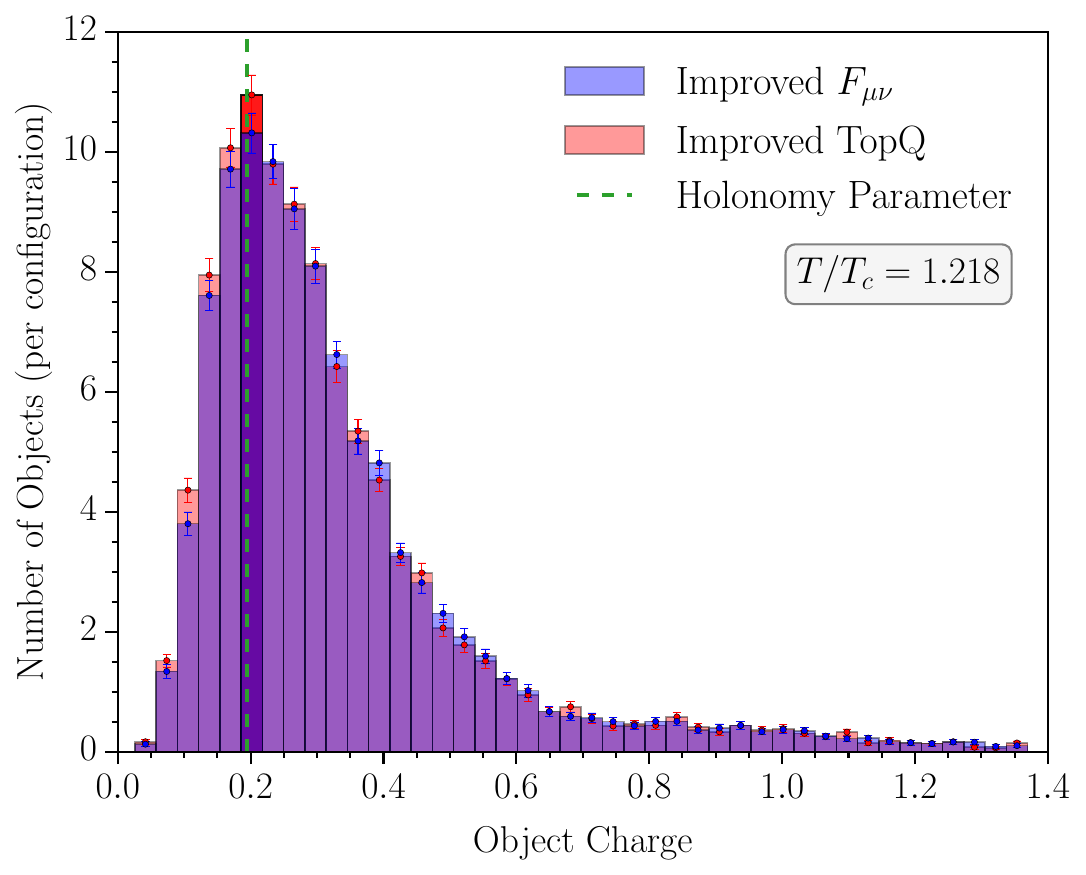}
	
	\vspace{0.5em}
	
	\includegraphics[width=0.49\linewidth]{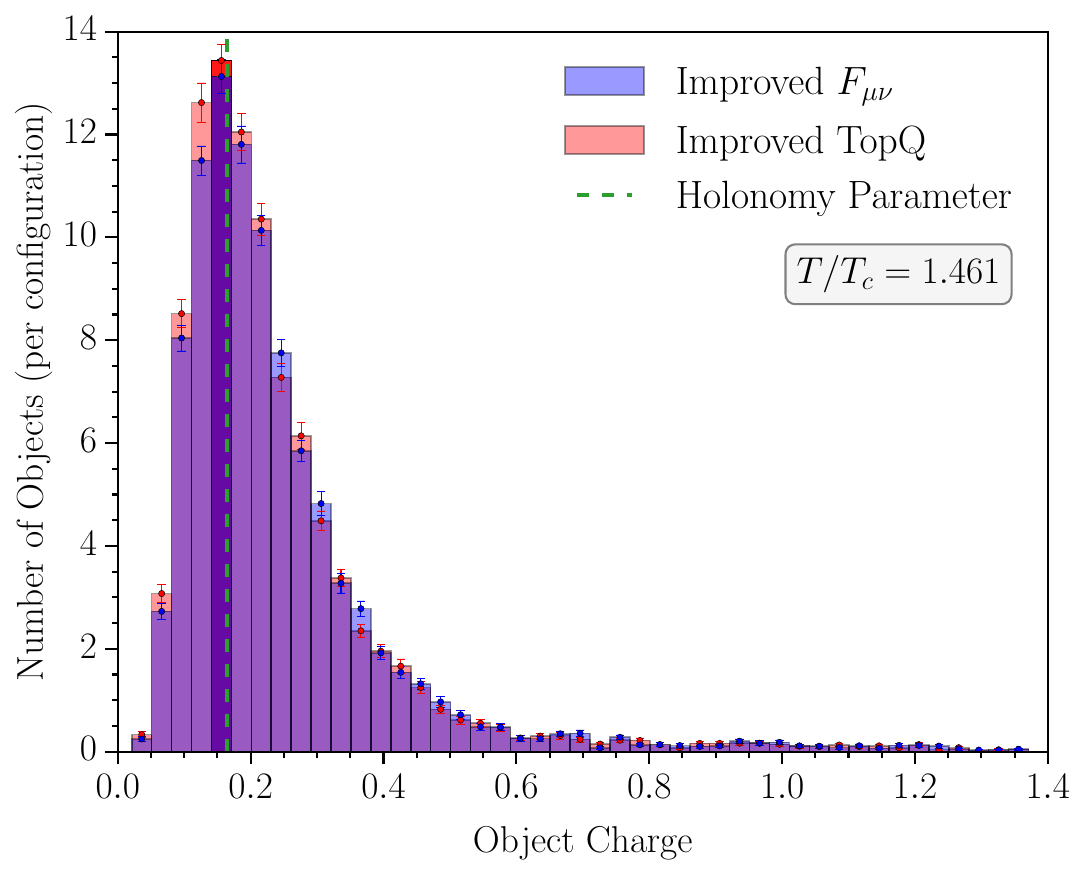}
	\hfill
	\includegraphics[width=0.49\linewidth]{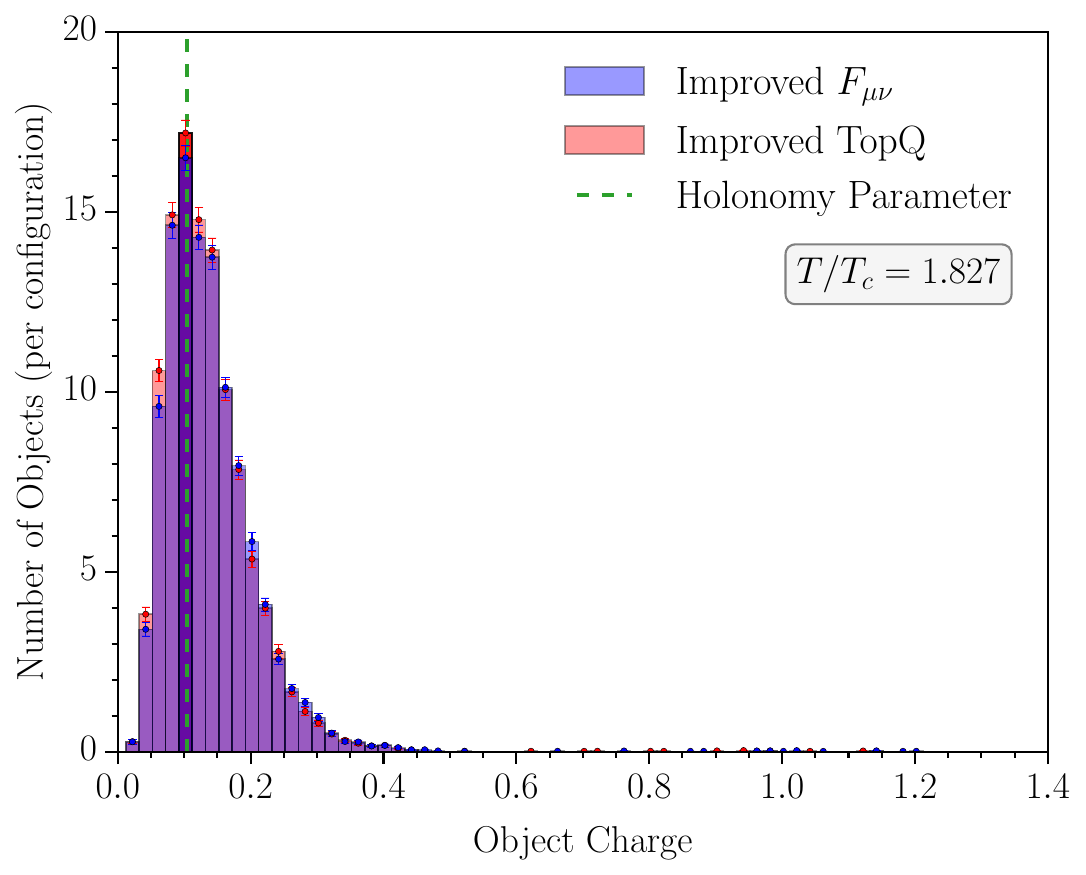}
	\caption{\label{fig:results} The result of our algorithm applied to each finite-temperature ensemble: $N_t = 64$ (\textbf{top left}), $12$ (\textbf{top right}), $8$ (\textbf{middle left}), $6$ (\textbf{middle right}), $5$ (\textbf{bottom left}) and $4$ (\textbf{bottom right}). The $\mathrm{SU}(3)$ free holonomy parameter described in text is shown with the dashed vertical line. The results are calculated after a flow time $\tau=1.45$.}
\end{figure}

For our analysis we use six pure-gauge ensembles with a spatial volume of $32^3$ and fixed lattice spacing of $a \simeq 0.1$\,fm, generated using HMC \cite{Duane:1987de, Kamleh:2004xk} with an Iwasaki renormalisation-group-improved action \cite{Iwasaki:1983iya, Iwasaki:1996sn}. The details are provided in Table~\ref{tab:ensembles}.
\begin{table}
	\centering
	\begin{tabular}{|r||rrr|rrr|}
		\hline
		$N_t$     &   64 &   12 &    8 &    6 &    5 &    4 \\
		\hline
		$T/T_c$   & 0.11 & 0.61 & 0.91 & 1.22 & 1.46 & 1.83 \\
		$T$ (MeV) &   31 &  164 &  247 &  329 &  395 &  493 \\
		\hline
	\end{tabular}
	\caption{\label{tab:ensembles} Details of the ensembles utilised in this work, including the number of temporal lattice sites $N_t$ and the corresponding temperatures, in terms of both the critical temperature $T_c$ and in MeV.}
\end{table}
The topological-charge histograms obtained from applying the algorithm to these ensembles with the prescribed smoothing are presented in Fig.~\ref{fig:results}.

Below the critical temperature, in the confined phase, we find that the histogram mode lies very close to $1/3$ for each ensemble in this regime. It is abundantly clear that the ground-state fields are dominated by fractionally charged topological excitations, as opposed to integer-charged instantons. The proximity of the mode to $1/3$ is consistent with the numerous constructions of fractional instantons that possess topological charge $\sim 1/N$ in $\mathrm{SU}(N)$ gauge theory \cite{tHooft:1981nnx, Gonzalez-Arroyo:1998hjb, Montero:1999by, Montero:2000pb, Gonzalez-Arroyo:2019wpu, DasilvaGolan:2022jlm, Anber:2023sjn, Nair:2022yqi}.

Moving above the critical temperature, into the deconfined phase, we find the histogram mode gradually shifts towards zero. The largest decrease is in our first ensemble above $T_c$, where the mode has fallen from $\simeq 1/3$ to $\simeq 0.2$. It thereafter steadily declines down to $\simeq 0.1$ at the highest temperature considered here. Thus, we see that the topological structure exhibits nontrivial evolution with temperature above the phase transition.

The behaviour of our histogram modes can be accurately predicted from the Polyakov loop, defined for each spatial position $\mathbf{x}$ as
\begin{equation}
	P(\mathbf{x}) = \mathcal{P} \exp\left(ig \int_0^{1/T} dx_4 \, A_4(x) \right) \in \mathrm{SU}(N) \,,
\end{equation}
where $\mathcal{P}$ is the path-ordering operator. Its trace $L(\mathbf{x}) = \frac{1}{3}\Tr P(\mathbf{x})$ acts as an order parameter for the phase transition in Yang-Mills theory, such that its vacuum expectation value $\langle L \rangle$ is zero in the confined phase but nonzero in the deconfined phase.

Of particular interest is the system's ``holonomy'', defined as the Polyakov loop at spatial infinity:\linebreak $\displaystyle{P_\infty = \lim_{|\mathbf{x}|\to\infty} P(\mathbf{x})}$. In $\mathrm{SU}(3)$, the holonomy can be parameterised by a single parameter $\nu$ as \cite{DeMartini:2021dfi}
\begin{equation} \label{eq:holonomy}
	\langle L_\infty \rangle = \frac{1}{3} + \frac{2}{3}\cos(2\pi\nu)
\end{equation}
Thus, in the confined phase where $\langle L_\infty\rangle = 0$, one finds $\nu = 1/3$. Conversely, in the deconfined phase where $\langle L_\infty\rangle > 0$, $\nu$ decreases from $1/3$ towards zero. This qualitatively agrees with the temperature dependence of our topological charge modes, and motivates a precise quantitative calculation of the free holonomy parameter $\nu$. The Polyakov loop on the lattice is calculated as
\begin{equation}
	P(\mathbf{x}) = \prod_{x_4=1}^{N_t} U_4(\mathbf{x},x_4) \,.
\end{equation}
Translational symmetry is exploited to compute $\langle L_\infty\rangle$ as the expectation value $\langle L\rangle$ of the spatially averaged Polyakov loop,
\begin{equation}
	L = \frac{1}{V} \sum_\mathbf{x} L(\mathbf{x}) \,.
\end{equation}

This process is complicated by the presence of a centre symmetry in the pure gauge theory. Below $T_c$, the Polyakov loop features a symmetry between the three centre phases in $\mathrm{SU}(3)$ \cite{Gattringer:2010ug, Danzer:2010ge, Stokes:2013oaa}. Above $T_c$, this symmetry is spontaneously broken with any one of the three centre phases becoming dominant. We overcome this by rotating the phase of $L$ by $\pm 2\pi i/3$ to bring the dominant phase of each configuration to a phase of zero. The ensemble average is subsequently taken to estimate $\langle L\rangle$.

The value of $\nu$ extracted from Eq.~(\ref{eq:holonomy}) is shown on the histograms in Fig.~\ref{fig:results} by a dashed vertical line, and is seen to be strongly coincident with the topological charge mode at all temperatures. This provides a very clear understanding of the topological structure of pure gauge theory at finite temperature---the dominant contribution at any given temperature is governed by the system's holonomy. This further establishes the Polyakov loop as vital in understanding the finite-temperature vacuum structure of pure gauge theory.

\section{Instanton-dyons}
One model that has the relationship between holonomy and topological structure built in is the instanton-dyon model. The ``instanton-dyons'' refer to the monopole constituents inside calorons, the finite-temperature generalisation of instantons \cite{Lee:1998vu, Lee:1998bb, Kraan:1998sn, Kraan:1998pm}. In general $\mathrm{SU}(N)$ gauge theory, the topological charges of the $N$ individual dyons are determined by the $N$ eigenvalues of the holonomy. In $\mathrm{SU}(3)$, these charges can be expressed in terms of the free holonomy parameter $\nu$ as \cite{DeMartini:2021dfi}
\begin{align}
	|Q_1| = \nu \,, && |Q_2| = \nu \,, && |Q_3| = 1 - 2\nu \,.
\end{align}
Two of the three dyons possess topological charges that match our histogram modes, as given by $\nu$.

The third dyon with topological charge $1 - 2\nu$ requires special attention. Below $T_c$, all three dyon charges are equal to $1/3$, in accordance with the histograms in this regime. Above $T_c$, the value of $1 - 2\nu$ increases towards $1$ as $\nu \to 0$, which at first glance is not necessarily reflected in our results. However, dyons are selfdual objects such that their actions and absolute topological charges are equal: $S_i = |Q_i|$. Accordingly, the third dyon with the larger topological charge, and hence action, will be exponentially suppressed in the partition function compared to the two dyons with topological charge $\nu$ \cite{DeMartini:2021dfi}. This accounts for the relative lack of objects with topological charge near $1 - 2\nu$ in our histograms above $T_c$. It is plausible that a small cluster of points near a charge of $1$ found in our highest-temperature histogram could be partly attributed to the presence of the third dyon type. As such, our results show a promising consistency with an instanton-dyon model for $\mathrm{SU}(3)$ vacuum structure at finite temperature.

\section{Conclusion}
We have presented a novel algorithm to directly estimate the topological charge contained within distinct topological objects in $\mathrm{SU}(3)$ lattice gauge theory. This is employed at finite temperature to build a comprehensive picture of how topological structure evolves with temperature. We find in the confined phase that the ground-state fields tend to be comprised of topological objects with fractional charge $\sim 1/3$, as defined by the modes of our topological-charge histograms. Above the phase transition, in the deconfined phase, this shifts towards zero.

The precise quantitative behaviour is accurately captured by the system's holonomy at each temperature. This provides a clear-cut relation between the Polyakov loop and topological structure, with our topological charge modes described by the free holonomy parameter $\nu$. Initial comparisons to the instanton-dyon model for finite-temperature vacuum structure are encouraging.

Extending this work to $\mathrm{SU}(N)$ for other values of $N$ is currently ongoing. It will be interesting to see whether the results naturally generalise to $N \neq 3$, where the expectation would be topological charges $\sim 1/N$ below the critical temperature. The first results presented here are promising.

\acknowledgments
This work was supported with supercomputing resources provided by the Phoenix High Performance Computing (HPC) service at the University of Adelaide. This research was undertaken with the assistance of resources and services from the National Computational Infrastructure (NCI), which is supported by the Australian Government. This research was supported by the Australian Research Council through Grant No.\ DP210103706. W.~K.\ was supported by the Pawsey Supercomputing Centre through the Pawsey Centre for Extreme Scale Readiness (PaCER) program.

\end{document}